\documentclass[onecolumn,showpacs,preprintnumbers,elsart]{revtex4}
\usepackage{amsmath}
\usepackage{color}
\usepackage{mathrsfs}
\usepackage{graphicx}
\usepackage{dcolumn}
\usepackage{bm}
\usepackage[center]{subfigure}

\begin{document}

\title{Double symmetry breaking of solitons in one-dimensional virtual
photonic crystals}
\author{Yongyao Li$^{1,4}$}
\email{yongyaoli@gmail.com}
\author{Boris A. Malomed$^{2,3}$}
\author{Mingneng Feng$^{1}$}
\author{Jianying Zhou$^{1}$}
\email{stszjy@mail.sysu.edu.cn}
\affiliation{$^{1}$State Key Laboratory of Optoelectronic Materials and Technologies,\\
Sun Yat-sen University, Guangzhou 510275, China\\
$^{2}$Department of Physical Electronics, School of Electrical
Engineering,
Faculty of Engineering, Tel Aviv University, Tel Aviv 69978, Israel\\
$^{3}$ICFO-Institut de Ciencies Fotoniques, Mediterranean Technology
Park,08860 Castelldefels (Barcelona), Spain\thanks{%
temporary Sabbatical address} \\
$^{4}$Department of Applied Physics, South China Agricultural
University, Guangzhou 510642, China}

\begin{abstract}
We demonstrate that spatial solitons undergo two consecutive spontaneous
symmetry breakings (SSBs), with the increase of the total power, in
nonlinear photonic crystals (PhCs) built as arrays of alternating linear and
nonlinear stripes, in the case when maxima of the effective refractive index
coincide with minima of the self-focusing coefficient, and vice versa, i.e.,
the corresponding linear and nonlinear periodic potentials are in
competition. This setting may be induced, as a \textit{virtual} PhC, by
means of the EIT (electromagnetically-induced-transparency) technique, in a
uniform optical medium. It may also be realized as a Bose-Einstein
condensate (BEC) subject to the action of combined periodic optical
potential and periodically modulated Feshbach resonance. The first SSB
happens at the center of a linear stripe, pushing a broad low-power soliton
into an adjacent nonlinear stripe and gradually suppressing side peaks in
the soliton's shape. Then, the soliton restores its symmetry, being pinned
to the midpoint of the nonlinear stripe. The second SSB occurs at higher
powers, pushing the narrow soliton off the center of the nonlinear channel,
while the soliton keeps its internal symmetry. The results are obtained by
means of numerical and analytical methods. They may be employed to control
switching of light beams by means of the varying power.
\end{abstract}

\pacs{42.65.Tg; 42.70.Qs; 05.45.Yv; 03.75.Lm}
\maketitle

%\email{yongyaoli@gmail.com}
%\email{stszjy@mail.sysu.edu.cn}

%\preprint{APS/123-QED}

% Force line breaks with \\



\section{Introduction}

It is commonly known that linear eigenstates supported by symmetric
potentials, in contexts such as quantum mechanics and photonic crystals
(PhCs), may be classified according to representations of the underlying
symmetry group \cite{LL}. The addition of the nonlinearity frequently gives
rise to the effect of the spontaneous symmetry breaking (SSB), i.e.,
reduction of the full symmetry group to its subgroups, in the generic
situation. The basic feature of the SSB is the transition from the symmetric
ground state (GS)\ to that which does not follow the symmetry of the
potential. The simplest manifestations of the SSB, which were predicted in
early works \cite{early}, and then in the model of nonlinear dual-core
optical fibers \cite{Snyder,fibers}, occur in settings based on symmetric
double-well potentials, or similarly designed nonlinear \textit{%
pseudopotentials} \cite{pseudo} (the latter name for effective potentials
induced by a spatially inhomogeneous nonlinearity is used in solid-state
physics \cite{Harrison}). In the quantum theory, the nonlinearity naturally
appears in the context of the mean-field description of Bose-Einstein
condensates (BECs), with the Schr\"{o}dinger equation replaced by the
Gross-Pitaevskii equation \cite{BEC}. Similarly, the self-focusing makes
PhCs a medium in which the linear symmetry competes with the nonlinearity,
in two-dimensional (2D) \cite{Valencia} and 1D \cite{Soukoulis}-\cite%
{defocusing} settings.

A natural extension of the SSB in the double-well potential is the
transition from symmetric to asymmetric solitons in the geometry which adds
a uniform transverse dimension to the potential, extending the potential
from the double well into a double trough. In this setting, the soliton
self-traps in the free direction due to the self-focusing nonlinearity. The
SSB effect in the solitons may be described by means of the effectively 1D
two-mode approximation, which is tantamount to the usual temporal-domain
model of dual-core optical fibers \cite{fibers}, and may also be applied to
the BEC loaded into a pair of parallel tunnel-coupled cigar-shaped traps
\cite{Arik}. In the application to the BEC, a more accurate description of
the symmetry-broken solitons was developed too, in the framework of the 2D
Gross-Pitaevskii equation, for the linear \cite{Marek} and nonlinear \cite%
{Hung} double-trough potentials.

PhC media are modeled by combinations of linear and nonlinear potentials,
which correspond to the alternation of material elements and voids in the
PhC structure \cite{Valencia}-\cite{defocusing}. A similar setting,
emulating the PhC, may be induced by means of the
electromagnetically-induced-transparency (EIT) technique in a uniform medium
\cite{we}. For BEC, a counterpart of the PhC may be generated as a
combination of the linear potential, induced by the optical lattice, and a
nonlinear pseudopotential imposed by a periodically patterned magnetic or
optical field modifying the local nonlinearity via the Feshbach-resonance
effect. Actually, the latter setting may be more general than the PhC, as
the so designed linear and nonlinear potentials may be created with
incommensurate periodicities \cite{HS}.

Solitons in periodic linear and nonlinear potentials have been studied
theoretically in many works, as reviewed in Refs. \cite{Barcelona} and \cite%
{Barcelona2}. In particular, specific types of \textit{gap solitons} were
predicted in 1D models of PhCs featuring the competition between the
periodically modulated refractive index and \emph{self-defocusing} material
nonlinearity \cite{defocusing}. Spatial optical solitons, supported by
effective linear potentials, were created in various experimental setups
\cite{experiment}.

Unlike the double-well settings, periodic potentials usually do not give
rise to SSB in solitons, although examples of asymmetric solitons were found
in some 1D models \cite{Kominis}. Indeed, 1D optical media built as periodic
alternations of self-focusing material elements and voids feature no
competition between the effective linear and nonlinear potentials, as minima
and maxima of both types of the potentials coincide, hence there is no drive
for the SSB in the medium (as mentioned above, the competition takes place
if the material nonlinearity is self-defocusing, but in that case the
corresponding gap solitons do not feature SSB either \cite{defocusing}).
Competing potentials leading to SSB effects might be possible if maxima of
the refractive index would correspond to minima of the self-focusing
nonlinearity. While this is impossible in usual PhCs, composed of material
stripes separated by voids, the effective potential structures induced by
EIT patterns in uniform media admit such a situation \cite{we} (as said
above, a similar setting is also possible in BEC \cite{HS,Barcelona2}). The
objective of this work is to study the SSB for spatial solitons in the
\textit{virtual} PhC of this type, following the gradual increase of the
total power of the soliton. It will be demonstrated that the symmetry
breaking happens twice in this setting, first to a low-power soliton
centered around a midpoint of a linear channel, and then to a high-power
beam situated at the center of the nonlinear stripe.

\begin{figure}[tbp]
%并排插入两个子图形
\centering {
\includegraphics[scale=0.3]{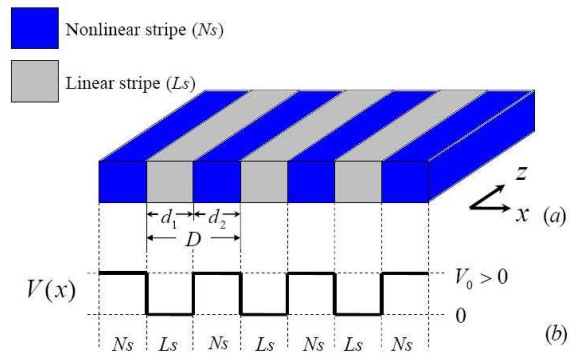}}
\caption{(Color online) (a) The scheme of the virtual photonic crystal with
period $D=d_1+d_2$. The blue (darker) and gray (lighter) slabs represent the
nonlinear and linear stripes,respectively. (b) The corresponding
Kronig-Penney modulation function.}
\label{fig_1}
\end{figure}

The setting under the consideration is displayed in Fig. \ref{fig_1}(a),
obeying the following version of the nonlinear Schr\"{o}dinger equation for
local amplitude $V(x,z)$, which is a function of propagation distance $z$
and transverse coordinate $x$:%
\begin{equation}
iu_{z}=-{(1/2)}u_{xx}+V(x)\left( 1-|u|^{2}\right) u,  \label{NLS0}
\end{equation}%
where the Kronig-Penney (KP) potential function, $V(x)$, is defined as per
Fig. \ref{fig_1}(b). We stress that the self-focusing sign of the
nonlinearity makes the model different from ones with the competition
between the linear and nonlinear potentials provided by the self-defocusing
\cite{defocusing}. We consider the version of the system with $%
d_{1}=d_{2}\equiv d$ in Fig. \ref{fig_1}(a).

The scaling is fixed by setting\ the modulation depth to be $V_{0}=0.02$,
which leaves half-period $d$ of the potential as a free parameter. In
Section 2, we present numerical results, at first, for $d=10$, which
illustrates a generic situation. Then, we demonstrate the results for other
values of the period, namely, $d=5,8,11,$ and $14$. In particular, it will
be demonstrated that the second SSB vanishes at large values of $d$. In
Section 3, we report analytical approximations,

\section{Solitons and their symmetry breakings (numerical results)}

\subsection{Results for $d=10$}

GS (ground-state) solutions to Eq. (\ref{NLS0}) with the above-mentioned
values of the parameters, $d_{1}=d_{2}=10$ and $V_{0}=0.02$, were found by
means of the imaginary-time-propagation method \cite{Chiofalo}.
Characteristic features of GS solitons, which distinguish them from bound
complexes of two or several solitary beams, is the presence of a pronounced
central peak, and the absence of nodes (zero crossings).

\begin{figure}[tbp]
%并排插入两个子图形
\centering
\subfigure[]{ \label{fig_2_a}
\includegraphics[scale=0.3]{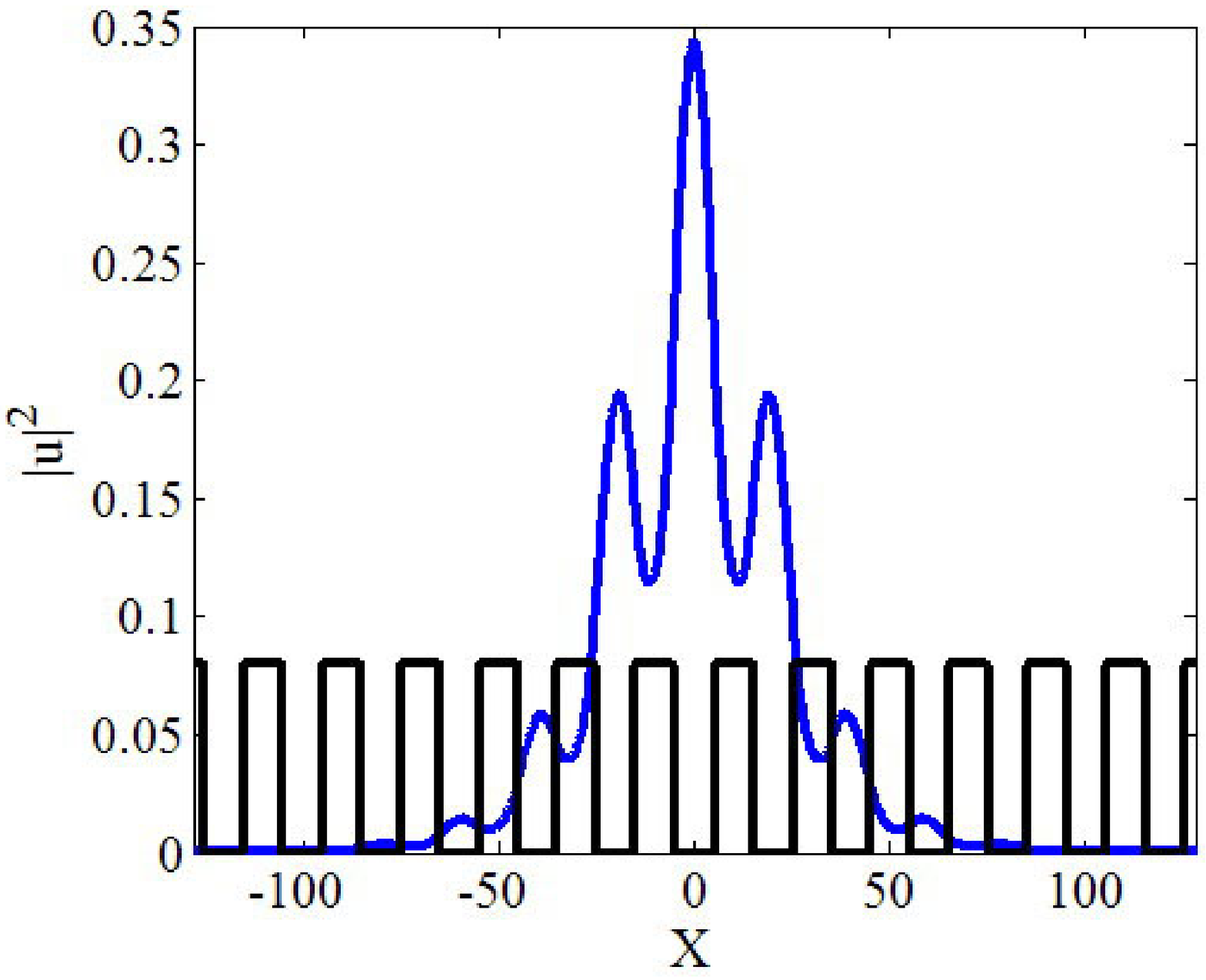}} \hspace{0.02in}
\subfigure[]{
\label{fig_2_b} \includegraphics[scale=0.3]{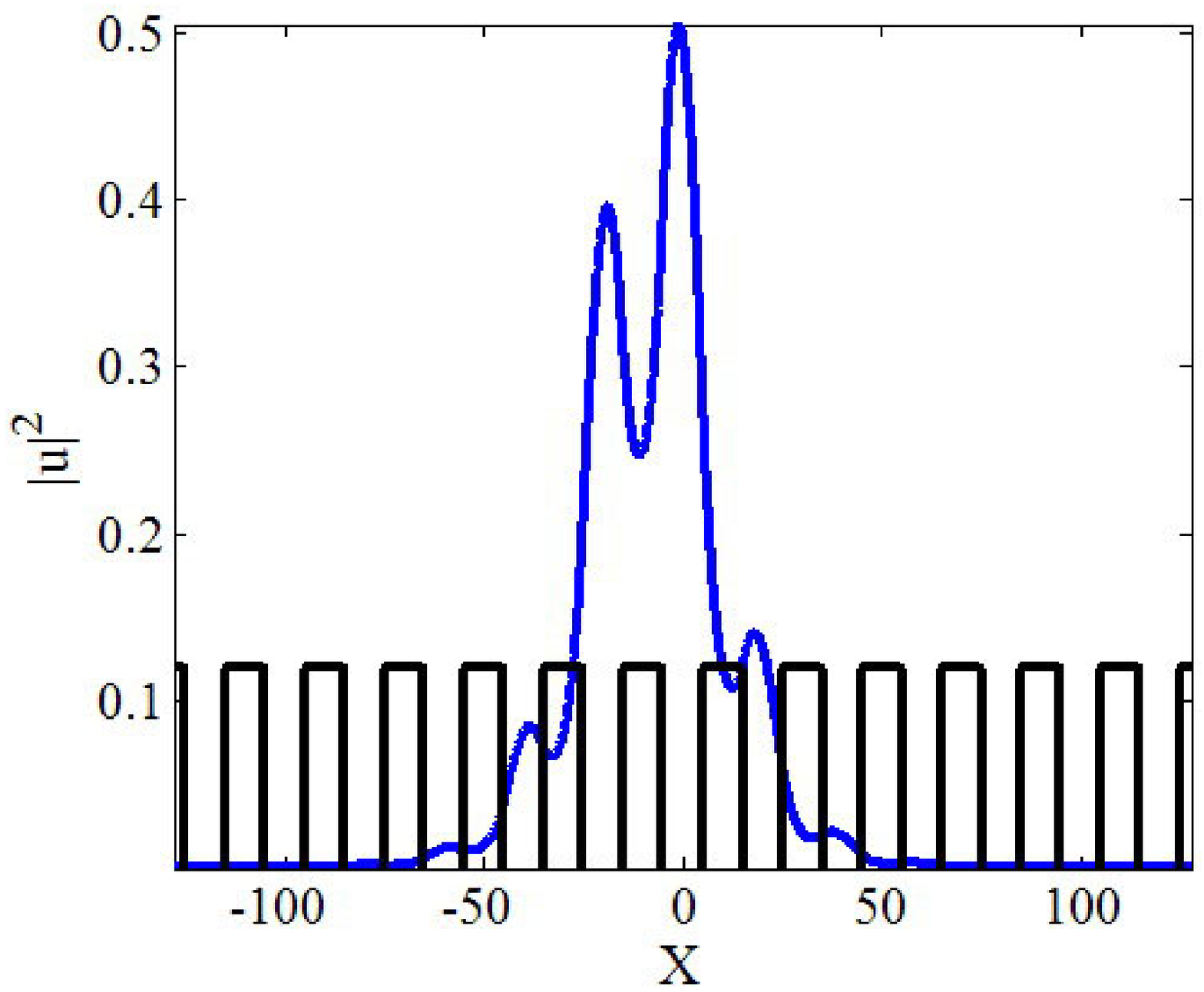}} \hspace{0.02in} %
\subfigure[]{ \label{fig_2_c} \includegraphics[scale=0.3]{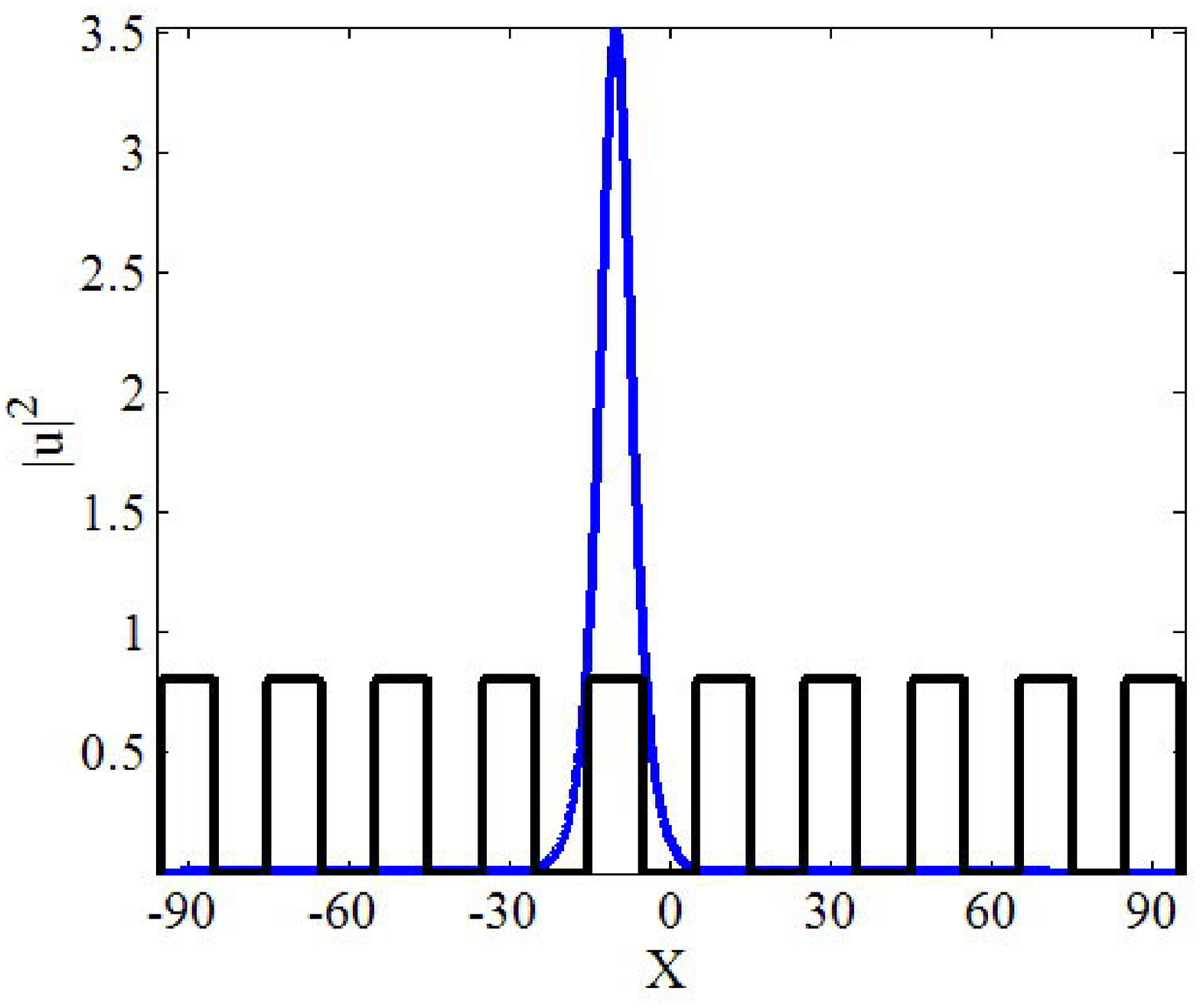}} \hspace{%
0.02in} \subfigure[]{ \label{fig_2_d}
\includegraphics[scale=0.3]{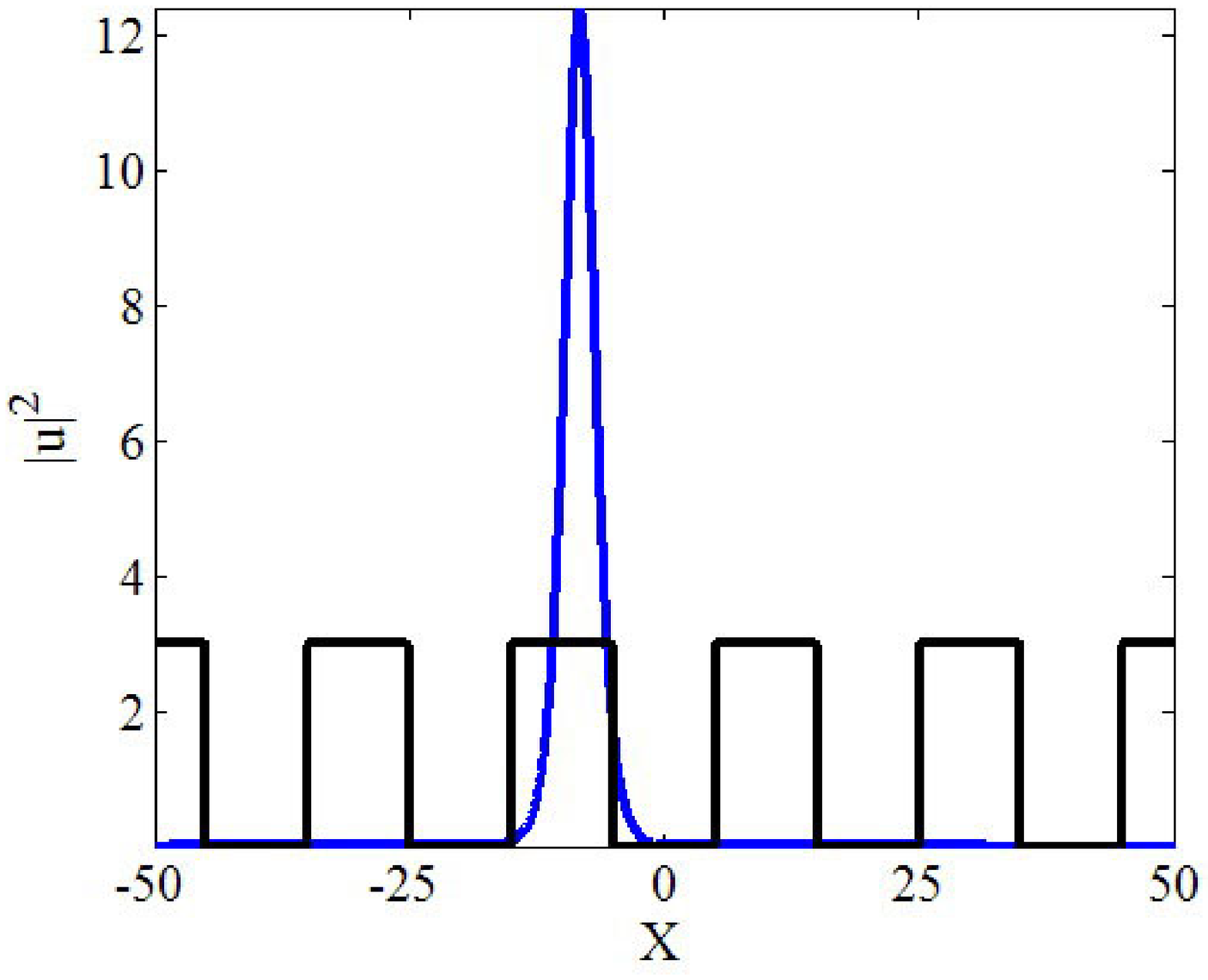}}
\caption{(Color online) Profiles of ground-state solitons found at the
following values of the total power: $P=12$ (a), $P=16$ (b), $P=29$ (c), and
$P=50$ (d).}
\label{fig_2}
\end{figure}

\begin{figure}[tbp]
%并排插入两个子图形
\centering
\subfigure[]{ \label{fig_4_a}
\includegraphics[scale=0.25]{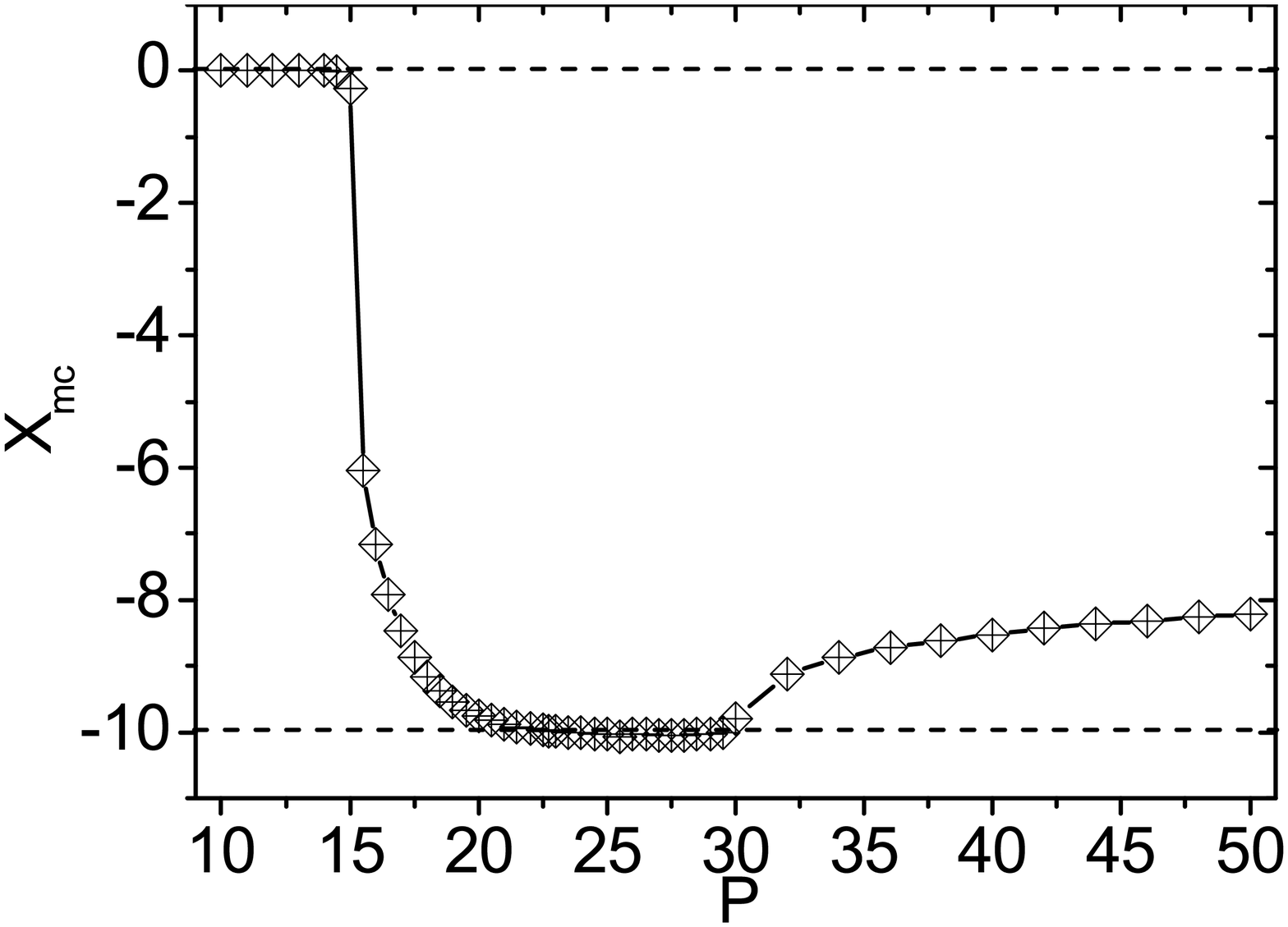}} \hspace{0.02in}
\subfigure[]{
\label{fig_4_b} \includegraphics[scale=0.25]{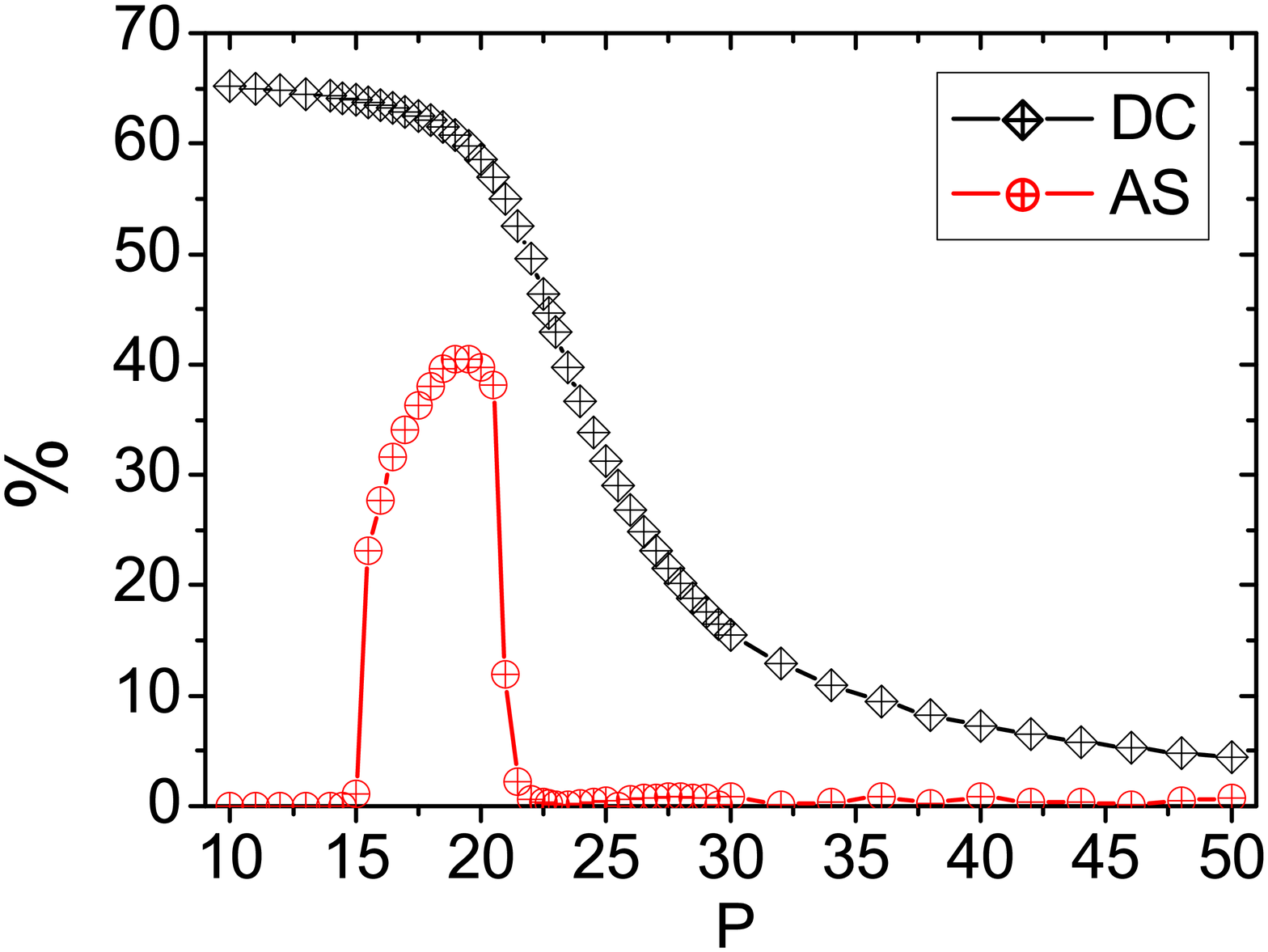}} \hspace{0.02in}
\subfigure[]{ \label{fig_4_c}
\includegraphics[scale=0.25]{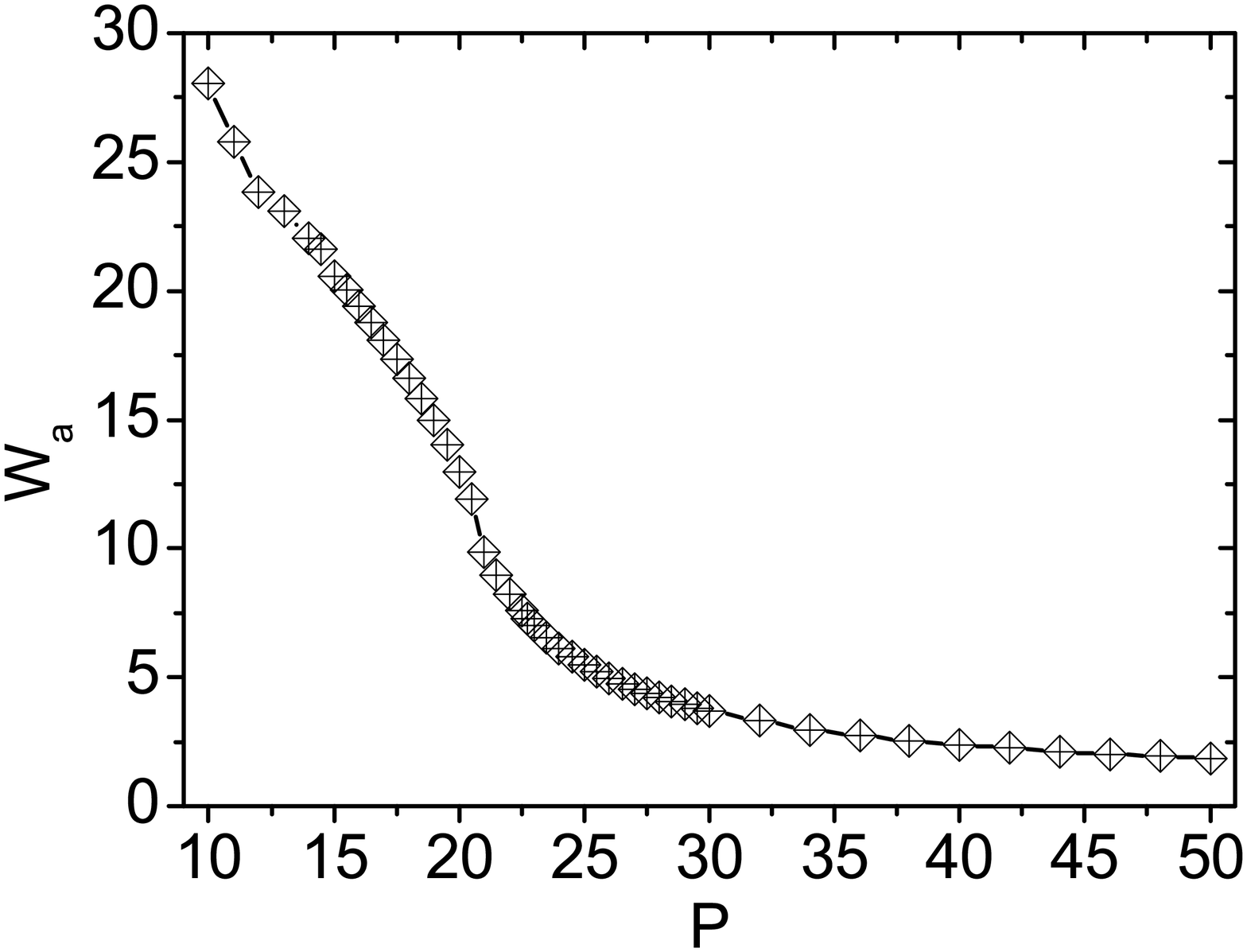}} \hspace{0.02in}
\subfigure[]{
\label{fig_4_d} \includegraphics[scale=0.25]{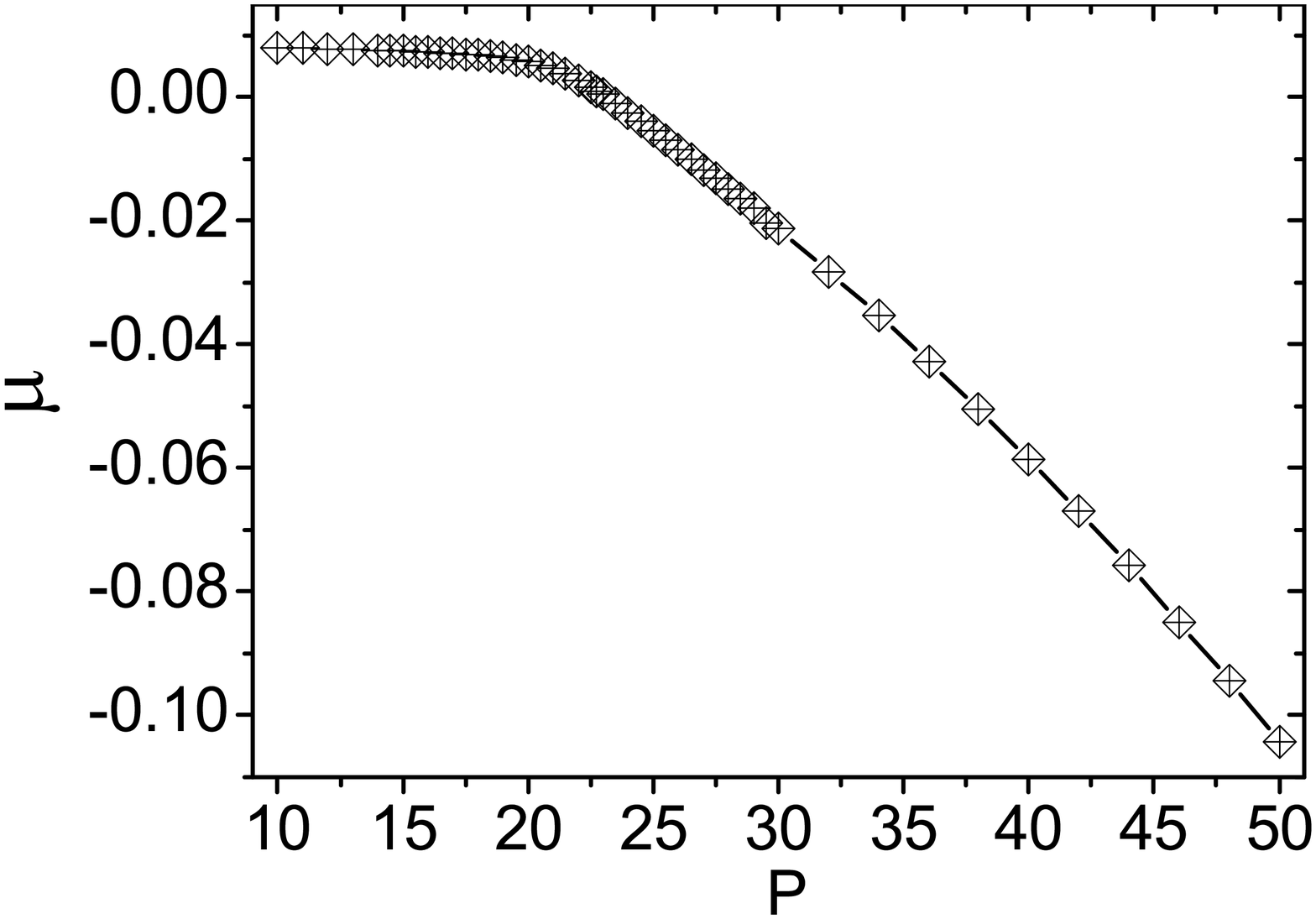}}
\caption{(Color online) (a) The coordinate of the soliton's center of mass
versus total power $P$. (b) The duty cycle (\textrm{DC}), showing the share
of the power trapped in linear stripes, and the soliton's asymmetry measure (%
\textrm{AS}), versus $P$. (c,d): The soliton's width ($W_{\mathrm{a}}$) and
propagation constant ($-\protect\mu $) as functions of $P$ (it can be
checked that all values of $\protect\mu $ belong to the semi-infinite gap,
in terms of the spectrum of the linearized system). Note that the $\protect%
\mu (P)$ dependence satisfies the Vakhitov-Kolokolov stability criterion
\protect\cite{VK}, $d\protect\mu /dP<0$.}
\label{fig_4}
\end{figure}

Figure {\ref{fig_2}} displays a characteristic set of GS profiles found at
different values of total power, which is defined as $P=\int_{-\infty
}^{+\infty }|u(x)|^{2}dx$. Representing the stationary solutions as $u\left(
x,z\right) =e^{-i\mu z}U(x)$, it is straightforward to check that wavenumber
$-\mu $ of all the GS solitons falls into the semi-infinite gap, in terms of
the spectrum generated by the linearized version of Eq. (\ref{NLS0}), see
Fig. \ref{fig_4}(d) below; this situation is natural for the system with the
self-focusing nonlinearity. Further, real-time simulations of Eq. (\ref{NLS0}%
) (not shown here) demonstrates that all the GS modes are stable against
perturbations. As seen from Fig. \ref{fig_4}(d), the stability of the GS
solitons is also supported by the \textit{Vakhitov-Kolokolov criterion} \cite%
{VK}, $d\mu /dP<0$.

At sufficiently small values of $P$ (in the weakly nonlinear regime), Fig. {%
\ref{fig_2}(a) demonstrates that }the central and side peaks of the GS
soliton are situated in linear channels, the soliton being symmetric about
its central peak, which exactly coincides with the midpoint of the
corresponding linear stripe. Although the light is chiefly guided by the
linear stripes in this regime, the soliton of course cannot exist without
the self-focusing nonlinearity, even if it is weak.

As seen in Fig. {\ref{fig_2}(b), }the first SSB event happens with the
increase of $P$, breaking the symmetry of the weakly-nonlinear GS soliton
and spontaneously shifting its central peak off the center of the linear
channel in which the peak is trapped. The symmetry of the side peaks gets
broken too, although they remain trapped in the linear channels. It is
relevant to stress that asymmetric GS solitons were not reported in
previously studied versions of nonlinear systems with KP potentials \cite%
{Kominis,defocusing,Merhasin}, although some exact asymmetric solutions for
non-fundamental solitons, which feature nodes, were found in Ref. \cite%
{Kominis}.

With the further increase of the power, the GS soliton undergoes strong
self-compression, which eliminates all side peaks, while the central peak
moves from the linear stripe into an adjacent nonlinear one, ending up in
the middle of the nonlinear stripe. A seen in Fig. {\ref{fig_2}(c),} in the
corresponding moderately nonlinear regime the single-peak soliton eventually
restores its symmetry about the midpoint of the nonlinear channel into which
it has shifted. Accordingly, light is guided by the nonlinear stripe in this
regime.

Finally, in the strongly nonlinear regime the \emph{second} SSB event
happens, as seen in Fig. \ref{fig_2}(d), where the narrow GS soliton
spontaneously shifts from the midpoint of the nonlinear stripe, although
staying in it. To the best of our knowledge, such a\emph{\ repeated} SSB of
solitons has not been reported in other models of nonlinear optics or BEC.
The second SSB seems a counter-intuitive effect, as one might
\textquotedblleft naively" expect that, with indefinite increase of total
power $P$, the narrow high-power soliton would be only stronger nested at
the center of the nonlinear stripe. Nevertheless, an explanation to this
effect is possible, as argued below.

To quantify the double SSB, we define the following characteristics of the
soliton, as functions of the total power, $P$: center-of-mass coordinate $x_{%
\mathrm{mc}}$, average width $W_{\mathrm{a}}$ , linear-stripe duty cycle (%
\textrm{DC}), the soliton's asymmetry measure (\textrm{AS}), and the
above-mentioned propagation constant, $-\mu $ (if the system is realized as
the BEC model, $\mu $ is the chemical potential):%
\begin{gather}
x_{\mathrm{mc}}=P^{-1}\int_{-\infty }^{+\infty }x|u|^{2}dx,  \notag \\
\mathrm{DC}=P^{-1}\int_{\mathrm{Ls}}|u|^{2}dx,~\mathrm{AS}%
=P^{-1}\int_{0}^{\infty }\left\vert \left[ u(x_{\max }-y)\right] ^{2}-\left[
u(x_{\max }+y)\right] ^{2}\right\vert dy,  \notag \\
W_{\mathrm{a}}^{2}=P^{-1}\int_{-\infty }^{+\infty }(x-x_{\mathrm{mc}%
})^{2}|u|^{2}dx,~\mu =P^{-1}\int_{-\infty }^{+\infty }u^{\ast }\mathbf{\hat{H%
}}udx,  \label{Symbols}
\end{gather}%
where $\int_{\mathrm{Ls}}$ stands for the integral taken over the linear
stripes, $x_{\max }$ is the location of the maximum value of $\left\vert
u(x)\right\vert $, and the Hamiltonian operator is $\mathbf{\hat{H}}={(1/2)}%
\partial _{xx}+V(x)[1-|u|^{2}]$. For $\mathrm{DC}>50\%$ ($<50\%$), light is
mainly guided by the linear (nonlinear) stripes. The strength of the SSB as
a whole is quantified by shift $x_{\mathrm{mc}}$, while $\mathrm{AS}$
quantifies the related inner symmetry breaking of the soliton.

The overall description of the double SSB is provided, in Fig. \ref{fig_4},
by plots showing the evolution of quantities (\ref{Symbols}) with the
increase of $P$. In panel (a), the dashed lines, $x=0$ and $x=-10$, mark the
positions of the two symmetric points in the KP potential, which correspond,
respectively, to midpoints of the linear and nonlinear stripes. The plots in
Figs. \ref{fig_4}(a,b) clearly demonstrate that the low-power GS soliton
remains symmetric, being centered at $x=0$, for $P<15$.

The first SSB event occurs at $P_{\mathrm{SSB}}^{(1)}\approx 15$. The
soliton becomes asymmetric, gradually moving from the midpoint of the linear
stripe ($x=0$) towards the center of the adjacent nonlinear one ($x=-10$),
in the interval of $15<P<22$. As seen in Fig. \ref{fig_4}(b), the soliton
attains the largest inner asymmetry degree, $\mathrm{AS}>40\%$, around $%
P\simeq 18$. Observe that the asymmetric soliton gradually loses its side
peaks, simplifying into the single-peak shape, as seen in Figs. \ref{fig_2}%
(a-c). The dependence $\mathrm{AS}(P)$ in Fig. \ref{fig_4}(b) shows that the
SSB occurring at $P=P_{\mathrm{SSB}}^{(1)}$ is of the \textit{supercritical
type}, i.e., it may be identified with a phase transition of the second kind
\cite{Landau}.

Next, the moderately high-power soliton remains completely symmetric,
centered at the midpoint of the nonlinear channel, at $22<P<30$. Note also
that the duty cycle falls to values $\mathrm{DC}<50\%$ at $P>22$, which
implies the switch from the quasi-linear guidance to that dominated by the
nonlinear pseudopotential.

The second SSB occurs at $P_{\mathrm{SSB}}^{(2)}\approx 30$. At $P>30$, the
high-power soliton gradually shifts from $x=-10$ towards the edge of the
nonlinear stripe, while keeping a virtually undisturbed symmetric shape,
with $\mathrm{AS}=0$, see Fig. \ref{fig_4}(b). This instance of the SSB may
also\ be considered as a phase transition of the second kind.

\subsection{Extension to other values of the potential's period}

\begin{figure}[tbp]
\centering
\subfigure[]{ \label{fig_5_a}
\includegraphics[scale=0.25]{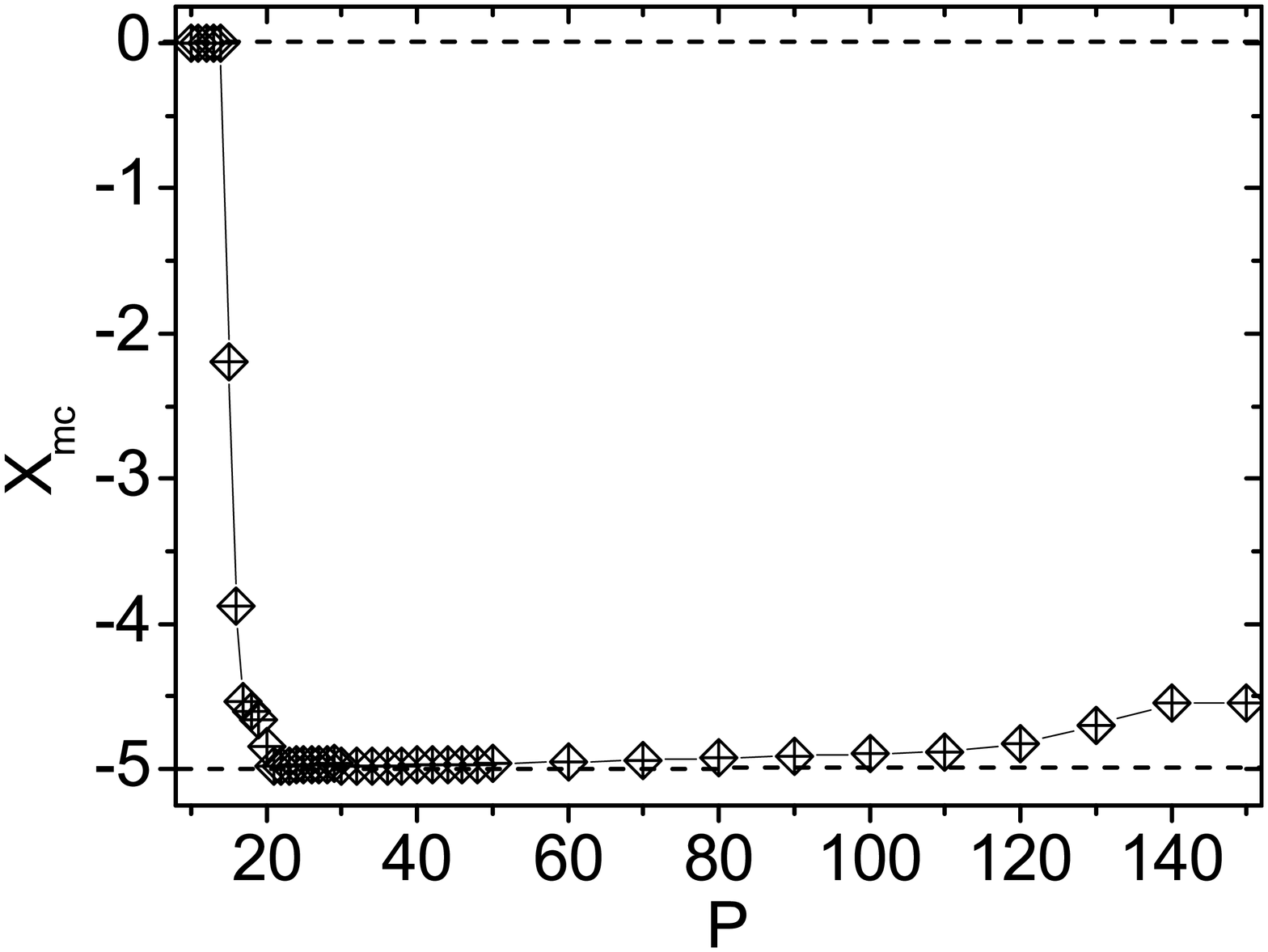}} \hspace{0.02in}
\subfigure[]{
\label{fig_5_b} \includegraphics[scale=0.25]{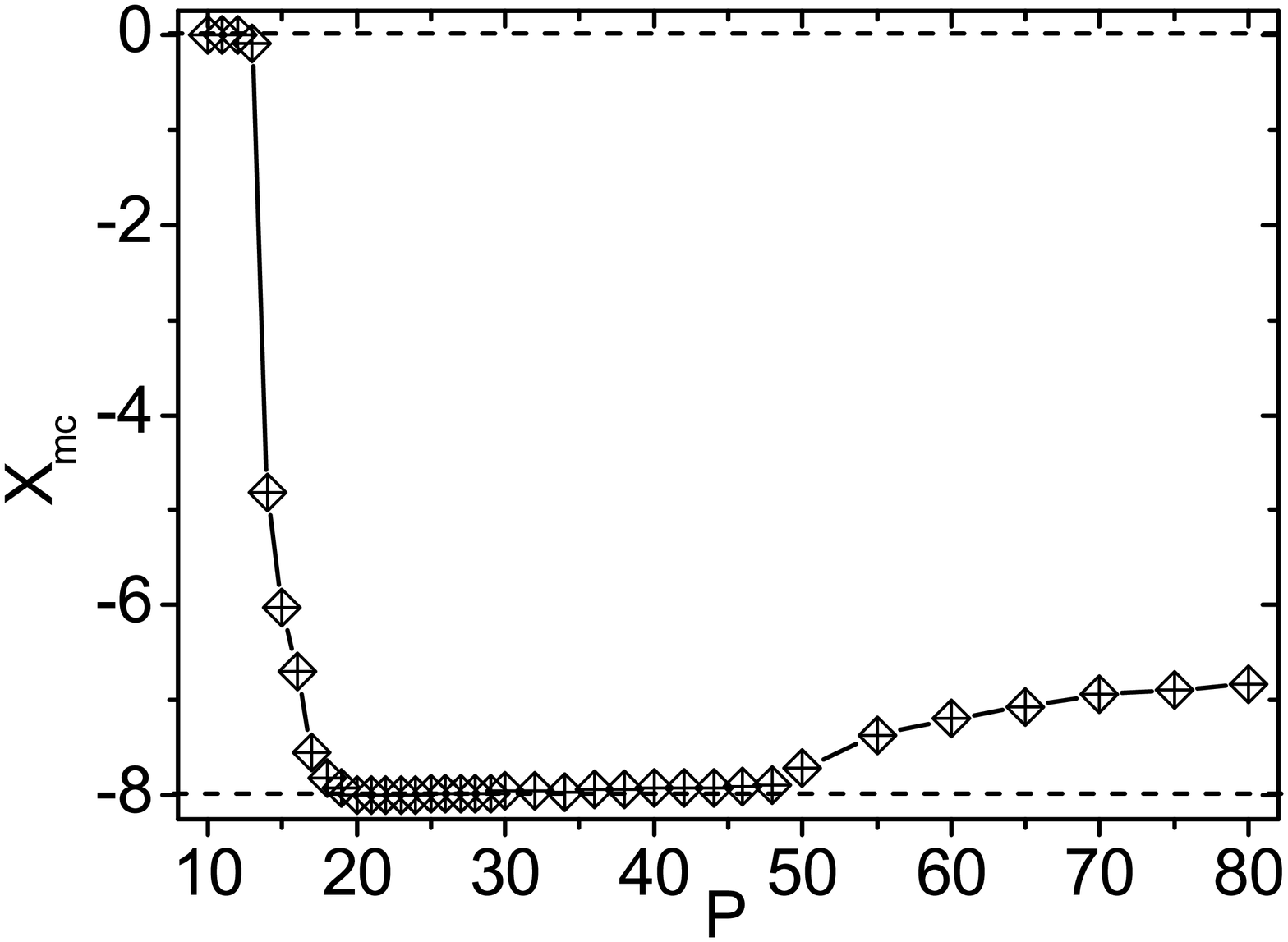}} \hspace{0.02in}
\subfigure[]{ \label{fig_5_c}
\includegraphics[scale=0.25]{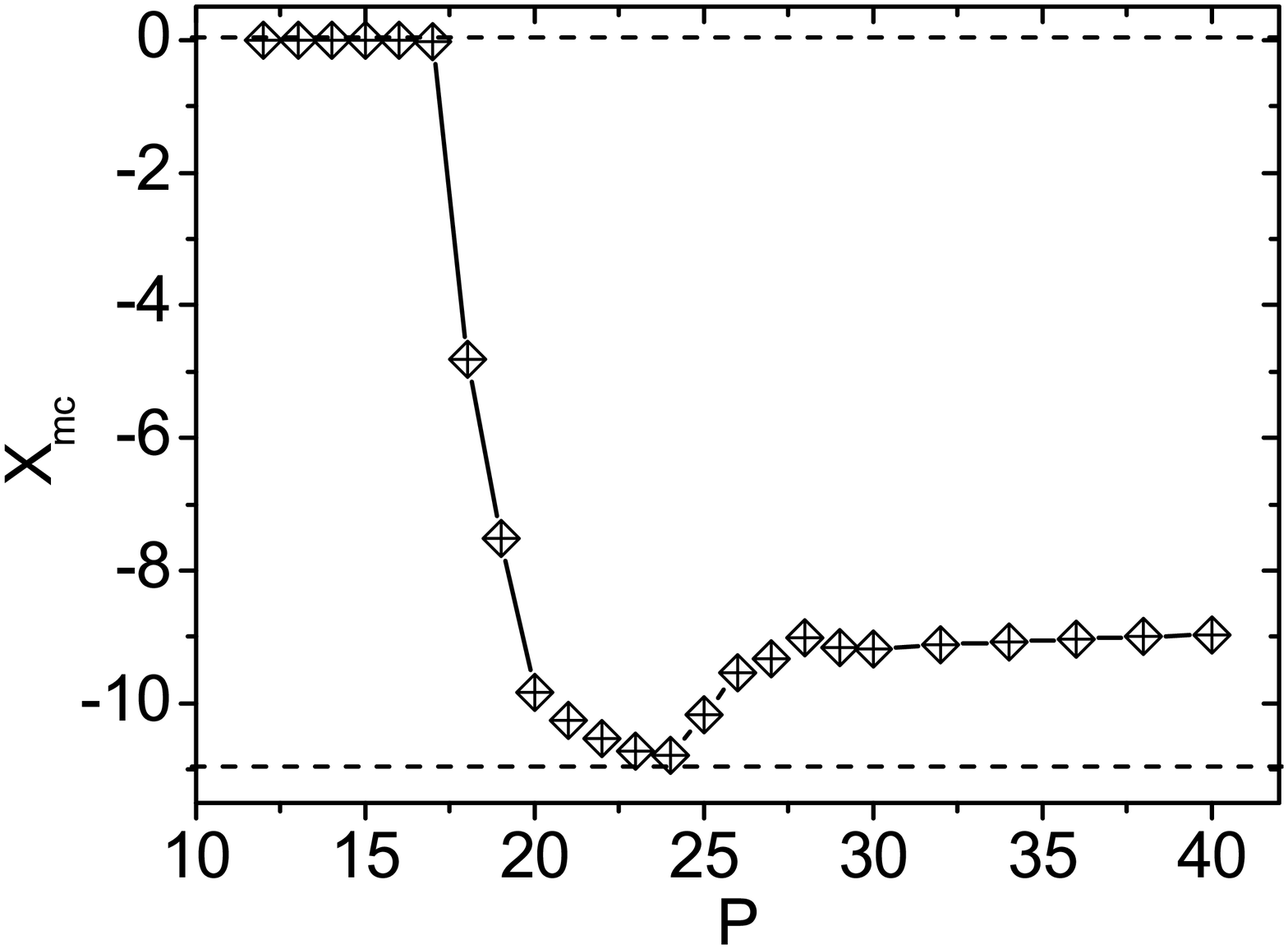}} \hspace{0.02in}
\subfigure[]{
\label{fig_5_d} \includegraphics[scale=0.25]{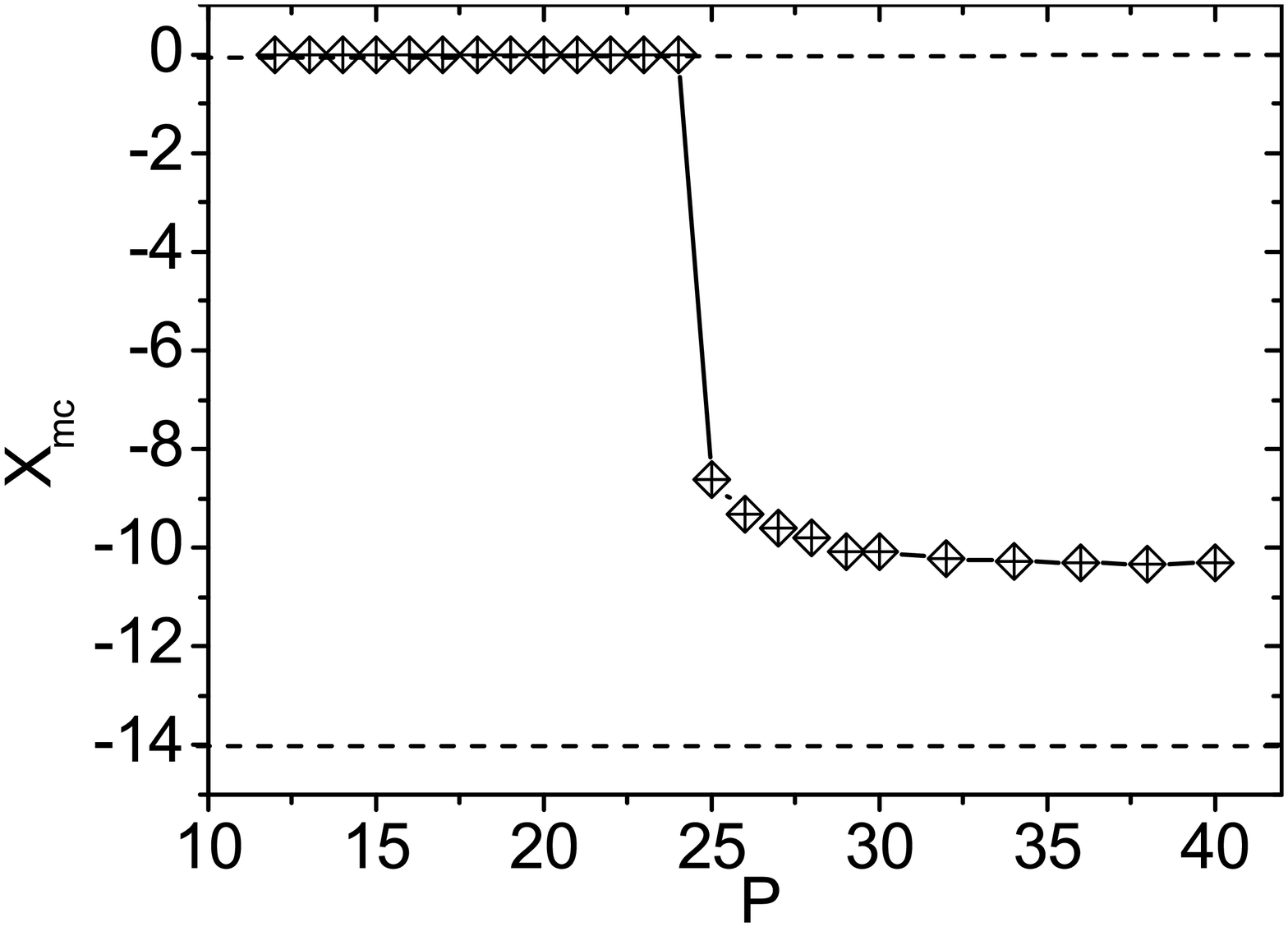}}
\caption{(Color online) $x_{\mathrm{mc}}(P)$ for different value of the
lattice period: (a) $d=5$, (b) $d=8$, (c) $d=11$, and (d) $d=14$}
\label{fig_5}
\end{figure}

To present the most general results, in Fig. \ref{fig_5} we report
dependence $x_{\mathrm{mc}}(P)$ for different values of $d$ (still with $%
d_{1}=d_{2}=d$), \textit{viz}., $d=5$, $8$, $11$, and $14$. We notice that,
when $d$ increases from $5$ in Fig. \ref{fig_5_a} to $11$ in Fig. \ref%
{fig_5_c}, the power at which the second SSB takes place decreases from $P_{%
\mathrm{SSB}}^{(2)}\approx 120$ to $\simeq 24$. This is explained by the
fact that the second SSB requires the average width of the soliton to be
much narrower than the width of the nonlinear stripe, hence, with smaller $d$%
, larger $P$ is needed to sustain the second SSB for a tighter localized
soliton. This argument also explains the shrinkage of the interval in which
the soliton is pinned to the symmetric position at the midpoint of the the
nonlinear stripe with the increase of $d$, as seen in Figs. \ref{fig_5_a}%
,(b) and (c). If $d$ keeps increasing, this interval eventually disappears
at $d>11$, as shown in Fig. \ref{fig_5_d}, pertaining to $d=14$.

\begin{figure}[tbp]
%并排插入两个子图形
\centering
\subfigure[]{ \label{fig_6_a}
\includegraphics[scale=0.25]{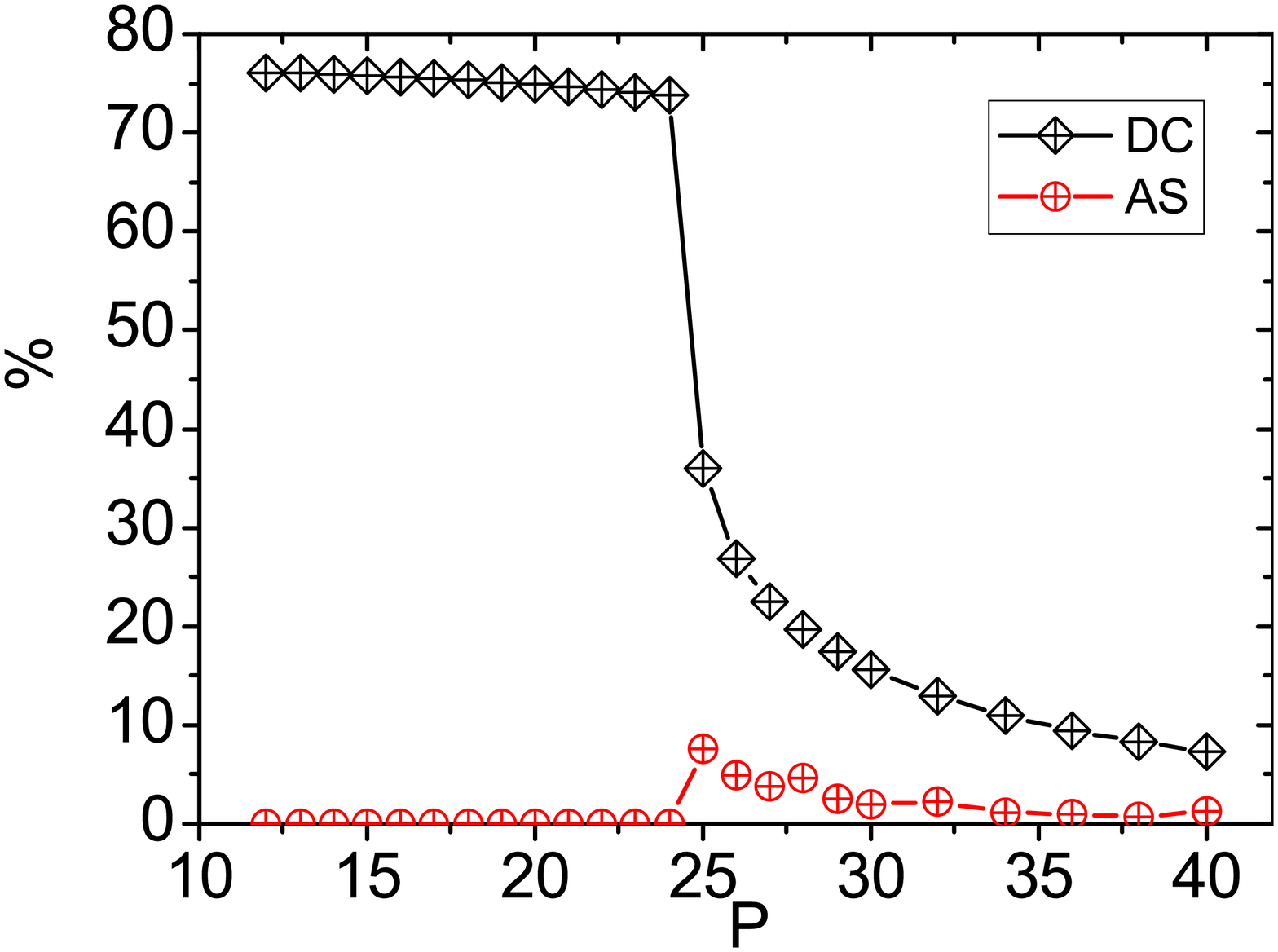}} \hspace{0.02in}
\subfigure[]{
\label{fig_6_b} \includegraphics[scale=0.25]{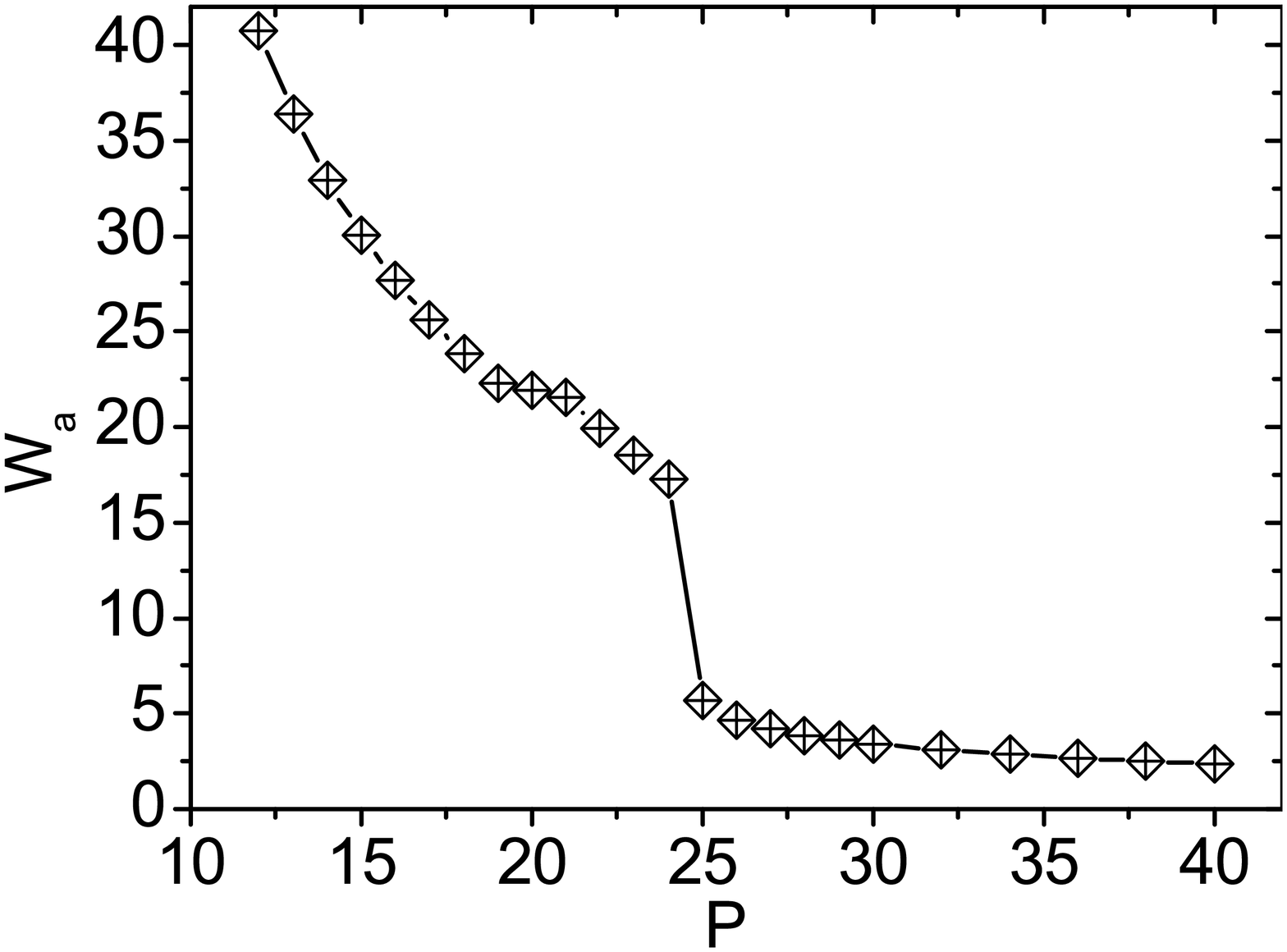}}
\caption{(Color online) (a) Dependences $\mathrm{DC}(P)$ and $\mathrm{AS}(P)$
for $d=14$. (b) $W_{\mathrm{a}}(P)$ for $d=14$.}
\label{fig_6}
\end{figure}

Further, in Fig. \ref{fig_6} we plot soliton characteristics $\mathrm{DC}$, $%
\mathrm{AS}$ and $W_{\mathrm{a}}$ for $d=14$. It exhibits a direct
transition from the original symmetric state to the final asymmetric state,
without the second SSB.

\section{Analytical considerations}

Both instances of the SSB revealed by the numerical findings can be
explained by means of analytical approximations. To address the first SSB,
which happens to the \emph{broad} low-power GS\ soliton, we replace the KP
modulation function with period $D\equiv 2d$ in Eq. (\ref{NLS0}), $V(x)$
(see Fig. \ref{fig_1}), by the combination of its mean value and the first
harmonic component, dropping higher-order harmonic components (cf. Ref. \cite%
{Wang}):%
\begin{equation}
\tilde{V}(x)=\left( V_{0}/2\right) [1-\left( 4/\pi \right) \cos \left( 2\pi
x/D\right) ],  \label{harmo}
\end{equation}%
Accordingly, Eq. (\ref{NLS0}) is replaced by
\begin{equation}
iu_{z}=-(1/2)u_{xx}+\tilde{V}(x)\left( 1-|u|^{2}\right) u.  \label{V0}
\end{equation}%
In the zero-order approximation, one may neglect the variable part of the
modulation function, whose period is much smaller than the width of soliton.
In this case, Eq. (\ref{V0}) gives rise to the obvious soliton solution,
\begin{equation}
u\left( x,z\right) =\sqrt{2/V_{0}}\eta e^{ikz}\mathrm{sech}\left( \eta
\left( x-\xi \right) \right) ,~k=\frac{1}{2}\left( \eta ^{2}-V_{0}\right) ,
\label{sol0}
\end{equation}%
with coordinate of the soliton's center $\xi $ and inverse width $\eta $.
The total power of the soliton is $P=4\eta /V_{0}$.

The Hamiltonian corresponding to Eq. (\ref{V0}) is%
\begin{equation}
H=\frac{1}{2}\int_{-\infty }^{+\infty }\left[ \left\vert u_{x}\right\vert
^{2}+\tilde{V}(x)\left[ 2|u|^{2}-|u|^{4}\right] \right] dx.  \label{H0}
\end{equation}%
As follows from here, the energy of the interaction of the broad soliton,
taken as per Eq. (\ref{sol0}) (which neglects the distortion of the
soliton's shape under the action of the potential) with the variable part of
modulation function (\ref{harmo}), expressed in terms of the soliton's
power, is%
\begin{gather}
U(\xi )\equiv -\left( V_{0}/\pi \right) \int_{-\infty }^{+\infty }\cos
\left( 2\pi x/D\right) \left[ 2|u(x)|^{2}-\left\vert u(x)\right\vert ^{4}%
\right] dx  \notag \\
=\frac{8\pi }{d}\left[ \sinh \left( \frac{4\pi ^{2}}{PV_{0}D}\right) \right]
^{-1}\left[ -1+\frac{2}{3V_{0}}\left( \left( \frac{\pi }{D}\right) ^{2}+%
\frac{P^{2}V_{0}^{2}}{16}\right) \right] \cos \left( \frac{2\pi \xi }{D}%
\right) .  \label{U0}
\end{gather}%
For small $P$ (low-power solitons), energy (\ref{U0}) gives rise to a \emph{%
local minimum} of the corresponding potential, i.e., a \emph{stable}
equilibrium position of the soliton at $\xi =0$, provided that
\begin{equation}
V_{0}>2\pi ^{2}/\left( 3D^{2}\right) .  \label{cond}
\end{equation}
Note that the parameter values adopted above to generate Figs. \ref{fig_1}-%
\ref{fig_4}, $D=20$ and $V_{0}=0.02$, satisfy this condition. Then, with the
increase of $P$, the equilibrium position predicted by potential (\ref{U0})
becomes unstable at

\begin{equation}
P>P_{\mathrm{SSB}}^{(1)}\equiv \left( 4/V_{0}\right) \sqrt{\left( 3/2\right)
V_{0}-\left( \pi /D\right) ^{2}}.  \label{eta}
\end{equation}%
At $P>P_{\mathrm{SSB}}^{(1)}$, the soliton moves away from $\xi =0$, i.e.,
this is the point of the first SSB. Note that, for $V_{0}=0.02$ and $D=20$,
expression (\ref{eta}) yields $P_{\mathrm{SSB}}^{(1)}\approx \allowbreak 14.6
$, which practically exactly coincides with the first SSB point identified
above from the numerical data, $P_{\mathrm{SSB}}^{(1)}\approx \allowbreak 15$%
. If $D$ is too small and does not satisfy condition (\ref{cond}) for given $%
V_{0}$, this means that the simplest approximation cannot be used, and
corrections to the average form of the soliton (\ref{sol0}) should be taken
into account, which is beyond the scope of the present analysis.

Proceeding to the analytical consideration of the second SSB, we rewrite Eq.
(\ref{NLS0}) as
\begin{equation}
iu_{z}=-(1/2)u_{yy}-V(y)\left( |u|^{2}-1\right) u,  \label{NLS}
\end{equation}%
where coordinate $y$ is defined so as to place the center of the nonlinear
stripe at $y=0$. Aiming to consider narrow solitons, trapped in the given
channel, which do not feel the presence of other nonlinear stripes, we
define the modulation function here so that $V(y)=V_{0}$ in the nonlinear
stripe, and $V(y)=0$ outside of it. Then, the narrow soliton with the center
located at point $y=\xi $ (generally, shifted off the center of the
nonlinear stripe) has the following form:%
\begin{equation}
u\left( z,x\right) =\frac{\eta e^{ikz}}{\sqrt{V_{0}}}\left\{
\begin{array}{c}
\mathrm{sech}\left( \eta \left( y-\xi \right) \right) ,~\mathrm{at}~~|y|~<%
\frac{d_{2}}{2}, \\
2\exp \left[ -\eta \left( \frac{d_{2}}{2}-\xi ~\mathrm{sgn}\left( y\right)
\right) -\sqrt{\eta ^{2}-2V_{0}}\left( |y|-\frac{d_{2}}{2}\right) \right] ~%
\mathrm{at}~|y|~>\frac{d_{2}}{2},%
\end{array}%
\right.  \label{sol}
\end{equation}%
where this time the inverse width of the soliton is assumed to be large, $%
\eta \gg 1/d_{2}$, $k$ is the same as in Eq. (\ref{sol0}), and the total
power of the narrow soliton is $P\approx 2\eta /V_{0}$.

The substitution of the wave field (\ref{sol}) into the Hamiltonian of Eq. (%
\ref{NLS}) yields the following effective potential of the interaction of
the soliton with the nonlinear stripe which holds it:%
\begin{equation}
U\left( \xi \right) =-2\eta \left[ \frac{\eta ^{2}}{V_{0}}\left( 1-\sqrt{1-%
\frac{2V_{0}}{\eta ^{2}}}\right) +2\right] e^{-\eta d_{2}}\cosh \left( 2\eta
\xi \right) +\frac{4\eta ^{3}}{V_{0}}e^{-2\eta d_{2}}\cosh \left( 4\eta \xi
\right) ,  \label{U}
\end{equation}%
cf. Eq. (\ref{U0}). As might be expected, the last term in this potential
tends to keep the soliton at the center ($\xi =0$), while the other terms
push it towards the edge of the nonlinear stripe. The equilibrium position
is defined by equation $dU/d\xi =0$. The substitution of potential (\ref{U})
into this equation yields two solutions: either $\xi =0$, which corresponds
to the soliton placed exactly at the center, and the off-center equilibrium,
determined by the following expression:%
\begin{equation}
\cosh \left( 2\eta \xi \right) =\frac{1}{8}\left[ \left( 1-\sqrt{1-\frac{%
2V_{0}}{\eta ^{2}}}\right) +\frac{2V_{0}}{\eta ^{2}}\right] e^{\eta d_{2}}.
\label{cosh}
\end{equation}%
Solution (\ref{cosh}) exists if it yields $\cosh \left( 2\eta \xi \right) >1$%
, i.e.,
\begin{equation}
e^{\eta d_{2}}>8\left[ \left( 1-\sqrt{1-2V_{0}/\eta ^{2}}\right)
+2V_{0}/\eta ^{2}\right] ^{-1}.  \label{exp}
\end{equation}%
Following the assumption that the soliton is narrow, we assume $2V_{0}/\eta
^{2}\ll 1$, hence Eqs. (\ref{cosh}) and (\ref{exp}) are simplified as
follows:%
\begin{gather}
\cosh \left( V_{0}P\xi \right) =3\left( 2V_{0}P^{2}\right) ^{-1}\exp \left(
V_{0}d_{2}P/2\right) ,~  \label{cosh2} \\
\exp \left( V_{0}d_{2}P/2\right) >(2/3)V_{0}P^{2}.  \label{exp2}
\end{gather}

The second SSB is realized, in the framework of the present approximation,
as the displacement of the soliton from $\xi =0$ to point (\ref{cosh}),
hence inequality (\ref{exp2}), if replaced by the respective equality,
offers a rough approximation for the second SSB point, $P_{\mathrm{SSB}%
}^{(2)}$ [\textquotedblleft rough" because it was derived taking into regard
exponentially small terms in Eq. (\ref{U}), which is a crude but meaningful
approximation \cite{expo}]. In particular, Eq. (\ref{exp2}) predicts that,
for $V_{0}=0.02$, the numerically found value reported above, $P_{\mathrm{SSB%
}}^{(2)}=30$, corresponds to $d_{2}\simeq 8.3$, which is not far from $%
d_{2}=10$ which was actually used. For $P\rightarrow \infty $, Eq. (\ref%
{cosh2}) yields $\xi \rightarrow \pm d_{2}/2$, i.e., within the framework of
the present approximation, the soliton moves to the edge of the nonlinear
stripe. On the other hand, for larger values of $d_{2}$ inequality (\ref%
{exp2}) always holds, which explains the disappearance of the second SSB in
the numerical picture displayed in Fig. \ref{fig_5}.

\section{Conclusion}

We have demonstrated the effect of the double SSB (spontaneous symmetry
breaking) for GS (ground-state) stable solitons in the 1D medium with the
competing periodic linear potential and its nonlinear counterpart
(pseudopotential) induced by a periodic modulation of the local
self-attraction coefficient. This medium may be realized as a virtual PhC
(photonic crystal) imprinted by means of the EIT technique into a uniform
optical medium, and also as the BEC setting using a combination of an
optical lattice and the spatially periodic modulation of the nonlinearity
via the Feshbach resonance. The two SSB events occur in the low- and
high-power regimes, pushing the soliton off the symmetric positions at the
center of the linear and nonlinear stripes, respectively. In the former
case, the SSB also affects the shape of the low-power soliton, making it
asymmetric and gradually stripping it of side peaks. In the latter case, the
narrow high-power soliton, while shifting off the midpoint of the nonlinear
channel, keeps the symmetric shape. At intermediate values of the power, the
soliton is completely symmetric, staying pinned at the center of the
nonlinear stripe. On the other hand, the increase of the period of the
potential structure leads to the direct transition from the original
symmetric state to the final asymmetric one, while the second SSB point
disappears. These results , which were obtained by means of systematic
numerical computations and explained with the help of analytical
approximations, suggest a possibility to control the switching of spatial
optical solitons in the virtual PhC by varying their power.

This work may be extended in other directions. It particular, a challenging
problem is to investigate similar settings and effects for 2D solitons, in
terms of PhCs and BEC\ alike.

Y.L. thanks Prof. X. Sun (Fudan University, Shanghai) for a useful
discussion. B.A.M. appreciates the hospitality of the State Key Laboratory
of Optoelectronic Materials and Technologies at the Sun Yat-sen University
(Guangzhou, China), and of the Department of Mechanical Engineering at the
Hong Kong University. This work was supported by the Chinese agencies NKBRSF
(grant No. G2010CB923204) and CNNSF(grant No. 10934011).

%\newpage %Just because of unusual number of tables stacked at end
%

\bibliographystyle{plain}
\bibliography{apssamp}

\begin{thebibliography}{Chiofalo(2000)}
\bibitem{LL} L. D. Landau and E. M. Lifshitz, \emph{Quantum Mechanics}
(Moscow: Nauka Publishers, 1974).

\bibitem{early} E. B. Davies, Comm. Math. Phys. \textbf{64}, 191 (1979); J.
C. Eilbeck, P. S. Lomdahl, and A. C. Scott, Physica D \textbf{16}, 318
(1985).

\bibitem{Snyder} A. W. Snyder, D. J. Mitchell, L. Poladian, D. R. Rowland,
and Y. Chen, J. Opt. Soc. Am. B \textbf{8}, 2101 (1991).

\bibitem{fibers} C. Par\'{e} and M. F\l orja\'{n}czyk, Phys. Rev. A \textbf{%
41}, 6287 (1990); A. I. Maimistov, Kvant. Elektron. \textbf{18}, 758 [Sov.
J. Quantum Electron. \textbf{21}, 687 (1991)]; N. Akhmediev and A.
Ankiewicz, Phys. Rev. Lett. \textbf{70}, 2395 (1993); P. L. Chu, B. A.
Malomed, and G. D. Peng, J. Opt. Soc. A B \textbf{10, }1379 (1993); B. A.
Malomed, in: Progr. Optics \textbf{43}, 71 (E. Wolf, editor: North Holland,
Amsterdam, 2002).

\bibitem{pseudo} L. C. Qian, M. L. Wall, S. Zhang, Z. Zhou, and H. Pu, Phys.
Rev. A \textbf{77}, 013611 (2008); T. Mayteevarunyoo, B. A. Malomed, and G.
Dong, \textit{ibid}. A \textbf{78}, 053601 (2008).

\bibitem{Harrison} W. A. Harrison, \emph{Pseudopotentials in the Theory of
Metals }(Benjamin: New York, 1966).

\bibitem{BEC} L. Pitaevskii and S. Stringari, \emph{Bose-Einstein
Condensation} (Clarendon Press: Oxford, 2003).

\bibitem{Valencia} P. Xie, Z.-Q. Zhang, and X. Zhang, Phys. Rev. E 67,
026607 (2003); A. Ferrando, M. Zacar\'{e}s, P. Fern\'{a}ndez de C\'{o}rdoba,
D. Binosi, and J. A. Monsoriu, Opt. Exp. \textbf{11}, 452 (2003); \textbf{12}%
, 817 (2004); J. R. Salgueiro, Y.i S. Kivshar, D. E. Pelinovsky, V. Sim\'{o}%
n, and H. Michinel, Stud. Appl. Math. \textbf{115}, 157 (2005); A. S.
Desyatnikov, N. Sagemerten, R. Fischer, B. Terhalle, D. Tr\"{a}ger, D. N.
Neshev, A. Dreischuh, C. Denz, W. Kr\'{o}likowski, and Y. S. Kivshar,
\textit{ibid}. \textbf{14}, 2851 (2006).

\bibitem{Soukoulis} Q. Li, C. T. Chan, K. M. Ho and C. M. Soukoulis, Phys.
Rev. B \textbf{53}, 15577 (1996); E. Lidorikis, Q. Li, and C. M. Soukoulis,
\textit{ibid}. \textbf{54}, 10249 (1996).

\bibitem{Wang} B. A. Malomed, Z. H. Wang, P. L. Chu, and G. D. Peng, J. Opt.
Soc. Am. B \textbf{16}, 1197 (1999).

\bibitem[Kominis(2006)]{Kominis} Y. Kominis, Phys. Rev. E. \textbf{73},
066619 (2006); Y. Kominis and K. Hizanidis, Opt. Exp. \textbf{16}, 12124
(2008).

\bibitem{defocusing} Y. Kominis and K. Hizanidis, Opt. Lett. \textbf{31},
2888 (2006); T. Mayteevarunyoo and B. A. Malomed, J. Opt. Soc. Am. B \textbf{%
25}, 1854 (2008).

\bibitem{Arik} A. Gubeskys and B. A. Malomed, Phys. Rev. A \textbf{75},
063602 (2007); A \textbf{76}, 043623 (2007).

\bibitem{Marek} M. Matuszewski, B. A. Malomed, and M. Trippenbach, Phys.
Rev. A \textbf{75}, 063621 (2007); M. Trippenbach, E. Infeld, J. Gocalek, M.
Matuszewski, M. Oberthaler, and B. A. Malomed, \textit{ibid}. A \textbf{78},
013603 (2008).

\bibitem{Hung} N. V. Hung, P. Zi\'{n}, M. Trippenbach, and B. A. Malomed,
Phys. Rev. E \textbf{82}, 046602 (2010).

\bibitem{we} C. Hang and V. V. Konotop, Phys. Rev. A \textbf{81}, 053849
(2010); Y. Li, B. A. Malomed, M. Feng, and J. Zhou, \textit{ibid.} \textbf{82%
}, 633813 (2010).

\bibitem{HS} H. Sakaguchi and B. A. Malomed, Phys. Rev. A \textbf{81},
013624 (2010).

\bibitem{Barcelona} Y. V. Kartashov, V. A. Vysloukh, and L. Torner, in:
Progr. Optics \textbf{52}, 63 (E. Wolf, editor: North Holland, Amsterdam,
2009).

\bibitem{Barcelona2} Y. V. Kartashov, B. A. Malomed, and L. Torner, \textit{%
Solitons in nonlinear lattices}, Rev. Mod. Phys., in press.

\bibitem{experiment} F. Lederer, G. I. Stegeman, D. N. Christodoulides, G.
Assanto, M. Segev, and Y. Silberberg, Phys. Rep. \textbf{463}, 1 (2008).

\bibitem[Chiofalo(2000)]{Chiofalo} M. L. Chiofalo, S. Succi, and M. P. Tosi,
Phys. Rev. E. \textbf{62}, 7438 (2000).

\bibitem[Merhasin(2005)]{Merhasin} I. M. Merhasin, B. V. Gisin, R. Driben,
and B. A. Malomed, Phys. Rev. E. \textbf{71}, 016613 (2005).

\bibitem{VK} M. Vakhitov and A. Kolokolov, Radiophys. Quantum. Electron.
\textbf{16}, 783 (1973); L. Berg\'{e}, Phys. Rep. \textbf{303}, 259 (1998).

\bibitem[Landau(2002)]{Landau} L. D. Landau and E. M. Lifshitz, \emph{%
Statistical Physics} (Moscow: Nauka Publishers, 1976).

\bibitem{expo} Yu. S. Kivshar and B. A. Malomed, Phys. Rev. Lett. \textbf{60}%
, 164 (1988); Rev. Mod. Phys. \textbf{61}, 763 (1989).
\end{thebibliography}
% Produces the bibliography via BibTeX.

\end{document}